\documentclass[prl,twocolumn,showpacs,superscriptaddress,amsmath,amssymb]{revtex4-1}
\usepackage{graphicx}
\usepackage{wasysym}

\newcommand{\tr}[1]{\textnormal{tr}{\left\{#1\right\}}}

\begin{document}

\title{Coarse-grained quantum state estimation for noisy measurements}

\author{Yong Siah Teo}
\affiliation{Department of Optics, Palacky University, 17. listopadu 12, 77146 Olomouc, Czech Republic}
\author{Jaroslav {\v R}eh{\'a}{\v c}ek}
\affiliation{Department of Optics, Palacky University, 17. listopadu 12, 77146 Olomouc, Czech Republic}
\author{Zden{\v e}k Hradil}
\affiliation{Department of Optics, Palacky University, 17. listopadu 12, 77146 Olomouc, Czech Republic}

\begin{abstract}
We introduce a straightforward numerical coarse-graining scheme to estimate quantum states for a set of noisy measurement outcomes, which are difficult to calibrate, that is based solely on the measurement data collected from these outcomes. This scheme involves the maximization of a weighted entropy function that is simple to implement and can readily be extended to any number of ill-calibrated noisy outcomes in a measurement set-up, thus offering practical applicability for general tomography experiments without additional knowledge or assumptions about the structures of the noisy outcomes. Simulation results for two-qubit quantum states show that coarse-graining can improve the tomographic efficiencies for noise levels ranging from low to moderately high values.
\end{abstract}

\date{\today}

\begin{widetext}
\maketitle
\end{widetext}

Quantum state tomography is one of the standard protocols for determining the integrity of the quantum state $\rho_\text{true}$ of a source. Typically, a set of probability operator measurement (POM) $\left\{\Pi_j\right\}$, with outcomes $\Pi_j\geq0$, is designed to measure a collection of quantum systems that are produced by the source. The measurement data collected are then the numbers of occurrences $n_j$ of all the outcomes $\Pi_j$. With these data, an estimator $\widehat{\rho}$ \cite{hat} for the source can be inferred and used as a book-keeping device to predict probabilities for future measurements or expectation values of any observable. Hypothetically, if the number of copies $N=\sum_jn_j$ measured approaches infinity, the estimator $\widehat{\rho}$ would essentially be the quantum state $\rho_\text{true}$ since the measured frequencies $f_j=n_j/N$ tend to the true probabilities $p_j$. In a realistic experimental scenario, however, the number $N$ is always finite and the resulting frequencies will not in general be physical probabilities. As such, we require more sophisticated methods of state estimation to ensure that the resulting estimator is positive.

Usually for a given tomography experiment, the POM that is used to perform the measurement is not exactly the intended POM of interest, owing to external random noise or systematic errors that perturb the measurement outcomes. One would need to calibrate these measurement outcomes before they can be used to reconstruct the unknown quantum state. Such calibrations are carried out by carefully performing separate experiments using well-defined probes to analyze the characteristics of the measurement \cite{det-calib}, which may be accompanied by distribution modeling for the external noise \cite{noisemod}. Ideally, if precise calibrations can be done for all the measurement outcomes, there exist quantum state estimation schemes available to reconstruct the state \cite{est-sch1,est-sch2,est-sch3}.

In this Letter, we are discussing state estimation for the case in which some, if not all, of the POM outcomes are difficult to calibrate. This can happen, for instance, if the noise perturbation evolves in such a way that there is no known distribution to describe such an evolution, or when hardware constraints simply render the task of calibration almost impossible. Under some assumptions about the noise and stability of the measurement outcomes, there exist self-calibrating techniques that simultaneously estimate the quantum state and certain aspects of the outcomes \cite{self-calib}. Using only the data $\left\{n_j\right\}$, our aim is to develop a straightforward numerical scheme to perform reliable quantum state estimation without any knowledge or assumptions about the noise distribution and the fine details of the noisy outcomes.

Throughout the discussion, we shall denote the well-calibrated POM outcomes as $\Pi^\text{(w)}_j$ and the ill-calibrated POM outcomes as $\Pi^\text{(i)}_k$. For the purpose of this discussion, we shall use the popular maximum-likelihood (ML) technique \cite{est-sch1} for state estimation. Suppose that out of a total of $M$ POM outcomes measured, only $M_1$ of them are well calibrated and $M_2=M-M_1$ outcomes are ill-calibrated and unknown. The likelihood functional for such a set of measurement outcomes $\sum_j\Pi^\text{(w)}_j+\sum_k\Pi^\text{(i)}_k\leq1$, given the data $\mathbb{D}=\left\{n^\text{(w)}_j;n^\text{(i)}_k\right\}$ obtained in a particular order, is defined as
\begin{align}
&\,\mathcal{L}\left(\mathbb{D};\rho\right)=\left[\prod^{M_1}_{j=1}\left(\dfrac{p^\text{(w)}_j}{\eta}\right)^{n^\text{(w)}_j}\right]\left[\prod^{M_2}_{k=1}\left(\dfrac{p^\text{(i)}_k}{\eta}\right)^{n^\text{(i)}_k}\right]\,,
\label{eq:like}
\end{align}
where $p^\text{(w)}_j=\tr{\rho\,\Pi^\text{(w)}_j}$, $p^\text{(i)}_k=\tr{\rho\,\Pi^\text{(i)}_k}$ and $\eta=\sum_jp^\text{(w)}_j+\sum_kp^\text{(i)}_k$. Let us also suppose that to every ill-calibrated outcome $\Pi^\text{(i)}_k$, there is a corresponding outcome $\Pi_k$ that one had intended to design and that the noisy outcome $\Pi^\text{(i)}_k$ is a result of noise perturbation on $\Pi_k$. Other than this, \emph{no other assumptions} are made regarding the actual measurement outcomes. Without loss of generality, we shall take the outcomes $\sum_j\Pi^\text{(w)}_j+\sum_k\Pi_k\leq1$ to be informationally complete, that is, these outcomes define a unique ML estimator. In principle, the subsequent arguments can be applied to informationally incomplete measurements.

Ideally, if we know precisely the identities of \emph{all} the outcomes, the actual ML estimator $\widehat\rho_\textsc{ml}$ can be reconstructed. Ruling out the calibration of the $\Pi^\text{(i)}_k$s as a viable option, one can conceive at least two strategies to go about estimating the state of the source in this situation. The first strategy is to simply take the data, neglect any noise in the system and reconstruct the ML estimator $\widehat{\rho}^{\text{ raw}}_\textsc{ml}$ by taking $\Pi^\text{(i)}_k=\Pi_k$ (\textbf{Strategy~1}), hoping that $\widehat{\rho}^{\text{ raw}}_\textsc{ml}$ is close to $\widehat\rho_\textsc{ml}$. If the actual $\Pi^\text{(i)}_k$s deviate from the respective $\Pi_k$s significantly, this strategy will not give an accurate estimator in general. The second strategy would be, since we are completely ignorant about the $\Pi^\text{(i)}_k$s, to discard all data obtained from measuring these ill-calibrated outcomes and use the rest to reconstruct the state (\textbf{Strategy~2}). Depending on the outcomes $\Pi^\text{(w)}_j$, the data may not be informationally complete. If so, there will in general be a convex set of estimators that give the same estimated probabilities. There are many ways of choosing a specific estimator from this set, none of which can systematically single out one that is close to $\widehat\rho_\textsc{ml}$ without additional information. Furthermore, if all the outcomes are not well calibrated, this strategy cannot be used.

\begin{figure*}[htp]
\centering
\includegraphics[width=1.7\columnwidth]{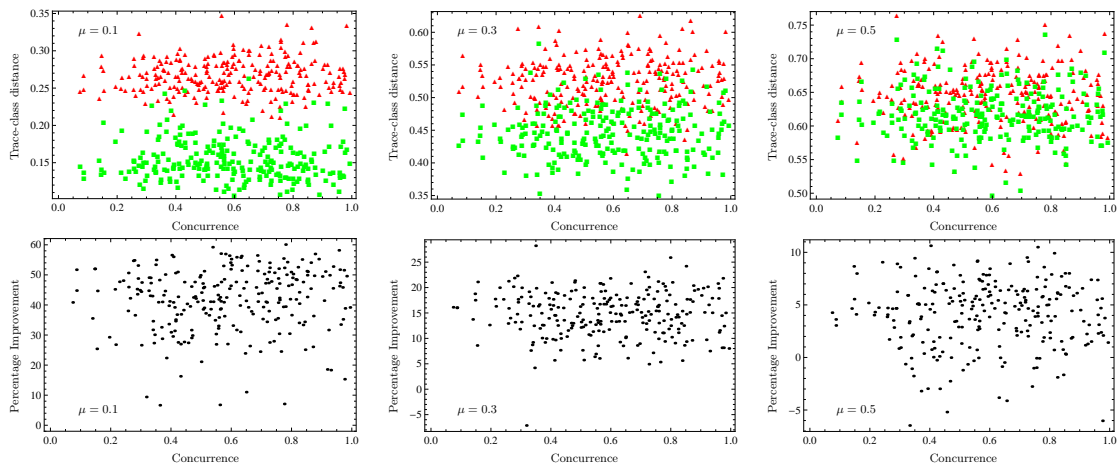}
\caption{Plots of the average trace-class distance over 20 experiments against the concurrence for 250 randomly generated pure states at noise levels $\mu=0.1$, 0.3 and 0.5. For moderately high values of $\mu$, the estimators $\widehat{\rho}^{\text{ cg}}_\textsc{ml}$ (Green~$\blacksquare$) are almost always closer to $\widehat{\rho}_\textsc{ml}$ than the estimators $\widehat{\rho}^{\text{ raw}}_\textsc{ml}$ (Red~$\blacktriangle$).}
\label{fig:trdist-spread}
\end{figure*}

We thus need to resort to another strategy and before we begin our modest attempt, we acknowledge that reasonable estimates for the $\Pi^\text{(i)}_k$s over the admissible space of POM outcomes are required to estimate $\widehat\rho_\textsc{ml}$ accurately. This is difficult to do given that the only available information is the set $\mathbb{D}$ and the complexity of this space. Instead, we take a different route and first carry out the replacement
\begin{equation}
\prod^{M_2}_{k=1}\left(\dfrac{p^\text{(i)}_k}{\eta}\right)^{n^\text{(i)}_k}\longrightarrow\prod^{M_2}_{k=1}\left(\dfrac{p_k}{\widetilde{\eta}}\right)^{n'_k}\,,
\label{eq:translation}
\end{equation}
with $\widetilde{\eta}=\sum_jp^\text{(w)}_j+\sum_kp_k$ and $p_k=\tr{\rho\,\Pi_k}$, for the term in Eq.~\eqref{eq:like} that is contributed by the ill-calibrated $\Pi^\text{(i)}_k$s, so that maximizing the resulting new likelihood functional will still give the same ML estimator $\widehat\rho_\textsc{ml}$ as maximizing the original one in Eq.~\eqref{eq:like}. In this way, we forgo the problem of estimating the $\Pi^\text{(i)}_k$s and turn to a fundamentally different problem of estimating the positive $n'_k$s, with the latter glossing over the operator details of the measurement and focusing on a reduced set of parameters. In this sense, this procedure results in a \emph{coarse-grained} parameter estimation. With only limited data, it is impossible to precisely obtain the correct $n'_k$s for which the replacement in Eq.~\eqref{eq:translation} leads to exactly $\widehat\rho_\textsc{ml}$. We will, next, estimate the $n'_k$s using the data $\left\{n^\text{(i)}_k\right\}$ and restrict the estimation using the relation
\begin{equation}
N=\sum^{M_1}_{j=1}n^\text{(w)}_j+\sum^{M_2}_{k=1}n^\text{(i)}_k=\sum^{M_1}_{j=1}n^\text{(w)}_j+\sum^{M_2}_{k=1}n'_k\,.
\end{equation}
This technique of coarse graining, which shall be our third strategy, attempts to perform an estimation on the set $\left\{n'_k\right\}$ with the raw data $\left\{n^\text{(i)}_k\right\}$ for the ill-calibrated outcomes and use the result, together with $\left\{n^\text{(w)}_j\right\}$, to reconstruct an estimator $\widehat{\rho}^\text{ cg}_\textsc{ml}$ (\textbf{Strategy~3}) using the intended outcomes $\Pi_k$. This strategy is meaningful only if the noise perturbation \emph{is not too large}.

We can now think of the raw data $n^\text{(i)}_k$ as a summary of our present subjective knowledge about the source and ill-calibrated outcomes. To estimate the parameters $n'_k$ for noise perturbation that is not too large, we will utilize a simple statistical inference method that takes this subjective knowledge into account. For small $N$, it can happen that $n^\text{(i)}_k=0$ for some $k$. In this case, we simply set the corresponding $n'_k=0$ since any nonzero values require justification the data cannot provide. Estimation shall then be performed on the $n'_k$s for which the $n^\text{(i)}_k$s are \emph{strictly positive}. By normalizing the two sets of parameters $\left\{n^\text{(i)}_k\right\}$ and $\left\{n'_k\right\}$ with $N_\text{ill}=\sum_kn^\text{(i)}_k$, we define the weighted entropy function
\begin{equation}
\mathcal{H}_{\left\{\nu^\text{(i)}_k\right\}}\left(\left\{\nu'_k\right\}\right)=-\sum^{M_{>0}}_{k=1}\nu^\text{(i)}_k\nu'_k\log\nu'_k
\label{eq:w-ent}
\end{equation}
to hold all information about the source and measurement, which is characterized by the parameters $\nu'_k=n'_k/N_\text{ill}$ to be estimated, with each outcome detection weighted by the respective subjective knowledge that is gained from the $M_{>0}\leq M_2$ positive $\nu^\text{(i)}_k=n^\text{(i)}_k/N_\text{ill}$. The relation between the concept of information and the weighted entropy function in Eq.~\eqref{eq:w-ent} was previously introduced and studied in Ref.~\cite{weight-ent}. The function $\mathcal{H}_{\left\{\nu^\text{(i)}_k\right\}}\left(\left\{\nu'_k\right\}\right)$ can also be regarded as a measure of the uncertainty of $\left\{\nu'_k\right\}$ given $\left\{\nu^\text{(i)}_k\right\}$. One approach of estimating the $n'_k$s, which we shall henceforth consider here, is to maximize the weighted entropy function with the constraint that $\sum_k\nu'_k=1$. The principle of such a weighted entropy maximization has already been used in the fields of Economics, Genetics and data pattern recognition \cite{weight-ent2}.
In effect, by maximizing the uncertainty with respect to our subjective knowledge about the ill-calibrated outcomes, we are searching for the parameters $n'_k$ that form the least-biased set with respect to the data $n^\text{(i)}_k$ \cite{est-sch3}. Based on this criterion, we deem this set as a conservative guess of the actual data that one would obtain if the intended outcomes $\Pi_k$ are used for measurement.

As the weighted entropy function $\mathcal{H}_{\left\{\nu^\text{(i)}_k\right\}}\left(\left\{\nu'_k\right\}\right)$ is a concave function in $\nu'_k$, it always gives a unique solution for its maximum and there is, hence, no ambiguity in the parameter estimation. Moreover, there is an arsenal of efficient nonlinear optimization methods for such simple convex functions. There are two other features in this coarse-grained \emph{maximum weighted entropy} (MWE) estimation scheme. Firstly, this scheme can be applied to any number of ill-calibrated outcomes $M_2$. This is particular useful if one believes that the entire set of measurement outcomes are noisy and not well calibrated, in which case he may choose $M_2=M$ to perform the coarse-grained MWE estimation. Secondly, this coarse-grained scheme does not rely on the details of the noisy outcomes. Such versatility permits its application to very general experimental situations without assuming any additional structures whatsoever about the noise and the ill-calibrated outcomes. From the MWE frequencies $\nu^{\textsc{mwe}}_k$ that maximize the weighted entropy function in Eq.~\eqref{eq:w-ent}, we obtain the set $\mathbb{D}_\textsc{mwe}=\left\{n^\text{(w)}_j;n^{\textsc{mwe}}_k\right\}$, with $n^{\textsc{mwe}}_k=N_\text{ill}\nu^{\textsc{mwe}}_k$, and use it to reconstruct an ML estimator by maximizing the new likelihood functional
\begin{equation}
\mathcal{L}\left(\mathbb{D}_\textsc{mwe};\rho\right)=\left[\prod^{M_1}_{j=1}\left(\dfrac{p^\text{(w)}_j}{\widetilde{\eta}}\right)^{n^\text{(w)}_j}\right]\left[\prod^{M_2}_{k=1}\left(\dfrac{p_k}{\widetilde{\eta}}\right)^{n^{\textsc{mwe}}_k}\right]\,.
\label{eq:mwe-like}
\end{equation}

\begin{figure*}[htp]
\centering
\includegraphics[width=1.8\columnwidth]{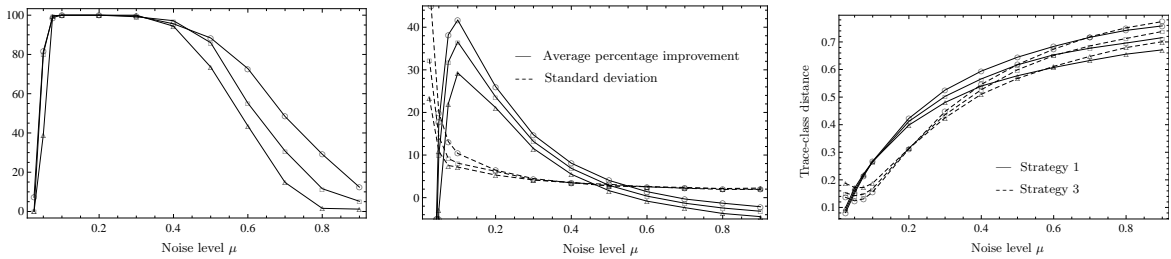}
\caption{Plots of percentage of total random states that respond better to Strategy~3 than Strategy~1~(Left), average and standard deviation of percentage improvement~(Center) and average trace-class distances with Strategy~1 and Strategy~3~(Right) for different noise levels $\mu$ and mixed quantum states of admixtures $\gamma=0\,(\ocircle)$, $0.1\,(\Box)$ and $0.2\,(\triangle)$. Nearly-pure mixed states yield comparable results. The intersections of the standard deviation curve and the average curve in (Center) for each $\gamma$ serve as a gauge for the performance range $\mathcal{R}_\gamma$ of Strategy~3. They are $\mathcal{R}_0\approx[0.05,0.55]$, $\mathcal{R}_{0.1}\approx[0.05,0.5]$ and $\mathcal{R}_{0.2}\approx[0.05,0.45]$.}
\label{fig:pure-mixed}
\end{figure*}

To put the third strategy to the test, we perform simulations involving noisy measurement outcomes. We first randomly generate a fixed set of $M=D^2$ rank-one positive operators $\widetilde{\Pi}_l$ such that $\sum_l\widetilde{\Pi}_l=1$, with $D$ being the dimension of the Hilbert space of interest and $l\in[1,M]$. Next, we select the first $M_1$ of these operators and define them to be the well-calibrated outcomes $\Pi^\text{(w)}_j=\mathcal{N}\widetilde{\Pi}_j$. From the rest of the $M_2$ operators which we define to be the intended outcomes $\Pi_k=\mathcal{N}\widetilde{\Pi}_{k+M_1}$, we construct the ill-calibrated outcomes as
\begin{equation}
\Pi^\text{(i)}_k=\mathcal{N}\left[\left(1-\mu\right)\widetilde{\Pi}_{k+M_1}+\mu\rho^\text{noise}_k\right]\,,
\end{equation}
where $\rho^\text{noise}_k$ is a randomly generated full-rank state with respect to the Hilbert-Schmidt measure and $0\leq\mu\leq1$ quantifies the noise level of the POM. The positive prefactor $\mathcal{N}$ ensures that the outcomes form a valid measurement. To model a varying noise perturbation, a different $\rho^\text{noise}_k$ is assigned to each experiment. We focus on two-qubit state estimation ($D=4$) on random two-qubit true quantum states $\rho_\text{true}$ generated with respect to the Hilbert-Schmidt measure. For each $\rho_\text{true}$, 20 experiments are simulated, where $N=8000$ copies of two-qubits are measured using the generated POM outcomes $\left\{\Pi^\text{(w)}_j;\Pi^\text{(i)}_k\right\}$ in each experiment.

For analysis, we suppose a typical situation in which $M_2=M$, so that Strategy~2 becomes impractical. We first compare the average performances of Strategy~1 and Strategy~3 for 250 random pure states. Figure~\ref{fig:trdist-spread} shows plots of the average trace-class distances between the estimator obtained with the first or the third strategy and $\widehat{\rho}_\textsc{ml}$, which is the ML estimator obtained with the actual POM measured. It is clear that when $\mu=0$, the trace-class distances between $\widehat{\rho}^\text{ raw}_\textsc{ml}$ and $\widehat{\rho}_\textsc{ml}$ will always be zero and Strategy~3 is not needed. For non-zero $\mu$, provided that $\mu$ is not too large, Strategy~3 turns out to be a better choice for state estimation than Strategy~1 on average, with percentage improvements that can exceed 50\%. For $\mu\gtrsim 0.55$, both strategies give very poor tomographic efficiencies since the data are too unreliable.

Next, it is interesting to look at the performance of Strategy~3 with mixed states. For this, we generate an ensemble of mixed states by taking each random pure state $\rho_\text{true}$ that is previously generated and admix to it the maximally-mixed state inasmuch as
\begin{equation}
\rho_\text{true}\longrightarrow(1-\gamma)\rho_\text{true}+\dfrac{\gamma}{4}\,,
\end{equation}
with $\gamma$ quantifying the amount of admixture. Figure~\ref{fig:pure-mixed} shows performance plots for three different values of admixtures. The drop in average performance of Strategy~3 with increasing $\gamma$ can be understood from the behavior that maximizing $\mathcal{H}_{\left\{\nu^\text{(i)}_k\right\}}\left(\left\{\nu'_k\right\}\right)$ as stated in Eq.~\eqref{eq:w-ent} tends to amplify the $\nu'_k$s with large weights and reduce those with small weights (refer to the second article in \cite{weight-ent}). As a result, this strategy can give very small $n'_k$s, especially when the relative weight differences are large. Maximizing $\mathcal{L}\left(\mathbb{D}_\textsc{mwe};\rho\right)$ then gives estimators that are nearly rank-deficient. Therefore, Strategy~3 is biased towards highly-pure estimators, which explains its overall effectiveness on pure true states for low to moderate $\mu$s. Without going into the details, we remark that mixed $\widehat{\rho}^\text{ cg}_\textsc{ml}$s of lower purity can be obtained by using adjustable weight factors $\left(\nu^\text{(i)}_k\right)^t$, with $t$ typically of values smaller than one, to reduce the relative weight differences appropriately. This has been tested to work in a preliminary investigation on experimental data for two-photon mixed states \cite{cqt-nus}. Coarse graining for mixed states and partially calibrated measurements will be reported in the future.

In conclusion, we have established a straightforward coarse-graining numerical method that can give accurate ML estimators for mixed quantum states that are nearly pure and ill-calibrated measurement outcomes of low and moderately high noise levels. This method employs the maximization of the weighted entropy in Eq.~\eqref{eq:w-ent} that requires no additional information about the the noise or the actual POM and can be applied to very general situations. In order to decide if coarse-grained MWE is suitable for a given set of experimental data from sufficiently large number of copies, one requires confidence that the unknown true state of the source is in the vicinity of the quantum state of interest, so that a credible comparison of the different strategies mentioned earlier can be made with respect to this state. Such an expectation is usually not too demanding as one usually has some prior knowledge about the source he is preparing, which can for instance be obtained through observations of physical aspects of the set-up. Finally, we would like to briefly mention that two other weighted entropy functions have been proposed respectively in Refs.~\cite{weight-ent3} and \cite{weight-ent4}. Experience shows that maximizing the weighted entropy function in Eq.~\eqref{eq:w-ent} gives more accurate ML estimators as the other two are less sensitive to the data and often yield rather flat distributions of parameters.

This work is co-financed by the European Social Fund and the state budget of the Czech Republic, project~No.~CZ.1.07/2.3.00/30.0004 (POST-UP), and supported by the Czech Technology Agency, project~No.~TE01020229.


\end{document}